\documentclass[aps,pre,twocolumn,superscriptaddress,,amsmath,amssymb,floatfix,ltxgridinfo]{revtex4-1} 
\usepackage{graphicx}
\usepackage{xcolor}
\usepackage{dcolumn}
\usepackage{bm}

\usepackage[utf8]{inputenc}
\usepackage[T1]{fontenc}
\usepackage{mathptmx}
\usepackage{etoolbox}
\usepackage{appendix}
\usepackage{lipsum}

\begin{document}

\title{Diffusive transport of a 2-D magnetized dusty plasma cloud}

\author{Aman Singh Katariya}
\email {amansk362@gmail.com} 
\affiliation{Department of Physics, Indian Institute of Technology Delhi, Hauz Khas, New Delhi 110016, India}

\author{Amita Das}
\email {amita@iitd.ac.in}
\affiliation{Department of Physics, Indian Institute of Technology Delhi, Hauz Khas, New Delhi 110016, India}

\author{Animesh Sharma }
\affiliation{Department of Physics, Indian Institute of Technology Delhi, Hauz Khas, New Delhi 110016, India}
		
\author{Bibhuti Bhushan Sahu}
\affiliation{Department of Energy Sciences and Engineering, Indian Institute of Technology Delhi, Hauz Khas, New Delhi 110016, India}

\begin{abstract}
Dusty plasma medium turns out to be an ideal system for studying the strongly coupled behavior of matter. The large size and slow response make their dynamics suitable to be captured through simple diagnostic tools. Furthermore, as the charge on individual particles is significantly higher than the electronic charge,  the interaction amongst them can be in a strong coupling regime even at room temperatures and normal densities. Such charged dust particles are often present in several industrial plasma-based processes and can have a detrimental influence. For instance, in magnetrons, the sputtering phenomena may be affected by the accumulation of charged impurity clusters.   The objective here is to understand the transport behavior of these particles in the presence of an externally applied magnetic field. For this purpose, Molecular Dynamics (MD) simulations are performed using an open-source large-scale atomic/molecular massively parallel simulator (LAMMPS) \cite{LAMMPS}. The dependence of the transport coefficient on the applied magnetic field and prevalent collisional processes has been discerned through simulations in detail.

\end{abstract}

\maketitle
\section{Introduction}
Dusty plasmas have been observed in a variety of contexts, ranging from astrophysical environments to laboratory industrial processes. For instance, they are involved in the etching process and thin film deposition.  They are also an ideal test bed for the fundamental understanding of strongly coupled many-body systems. The demonstration in the  1990s, of the formation of dust plasma crystals \cite{crystal1,crystal2,crystal3,crystal4} bears testimony of its strongly coupled characteristics. In many experiments, dust is introduced deliberately, but sometimes dust can accumulate in the experimental setup by itself and contaminate the system (e.g. in the context of industrial plasmas which are used for the deposition of thin films). The formation of dust and its dynamics inside the plasma has been studied quite extensively in the last decade \cite{study1,thomaslynch,karasev,jaiswal}, both numerically and experimentally. The large inertia associated with the dust particles and their high charge enables them to elicit strong coupling behavior at normal temperatures and densities.

In the last few decades, extensive studies \cite{Rao2000ElectrostaticMI, tsytovich2007development, DustEMmodes, ratynskaia2006electrostatic, kaw1998low, kumar2019coupling, kumar2018spiral, Nakamurashocks,merlino1998laboratory, das2014suppression, veeresha2005rayleigh,dharodi2022kelvin, dolai2022kelvin, Prerana_2014_Jeans, dolai2020effects, sharma2021effect, das2014collective, das2014exact, das2010nonlin, bandyo2008nonlin,  deshwal2022chaotic, maity2018interplay, maity2022parametric, maity2020dynamical, maity2019molecular,yadav2023structure} have been performed on strongly coupled  plasma behaviour. This has led to the observation and understanding of a variety of physical phenomena. For instance, the altered behavior of electrostatic modes has been investigated \cite{Rao2000ElectrostaticMI, tsytovich2007development, DustEMmodes, ratynskaia2006electrostatic, kaw1998low, kumar2019coupling, kumar2018spiral, Nakamurashocks}. Various instabilities such as Rayleigh-Taylor instability \cite{merlino1998laboratory, das2014suppression, veeresha2005rayleigh}, Helmholtz instability \cite{dharodi2022kelvin, dolai2022kelvin},  and Jeans instability \cite{Prerana_2014_Jeans, dolai2020effects, sharma2021effect} have been studied in such plasmas. There are also studies on nonlinear dynamics associated with such systems \cite{das2014collective, das2014exact, das2010nonlin, bandyo2008nonlin,  deshwal2022chaotic}. The formation of ordered structures in such a medium has also captured the research interest\cite{maity2018interplay, maity2022parametric, maity2020dynamical, maity2019molecular,yadav2023structure}.  

The size of the dust particle typically varies from nanometer to micrometer scale. The size determines how much charge can accumulate on its surface.  The processes which lead to charges adhering to its surface involve phenomena such as electron attachment from background plasma, photo-electron emission, secondary electron emission, thermionic emission, etc. \cite{pkshuklareview}. Most of these processes are surface-related phenomena.  The typical charges that can accumulate on a nano-sized dust particle can be of the order of  $\sim$ a few tens of electrons to thousands of electrons \cite{tadsen2015self, staps2021situ}. For charge particles of this size thus an external magnetic field of the order of $\sim$ 10 to 50 mT \cite{gudmundsson2020physics} (typically employed in the context of magnetron devices) can easily elicit a magnetized response. The characteristic behavior of dust particle transport in such a situation might, therefore, be significantly different from their unmagnetized behavior.  The strongly coupled characteristic interaction of these particles opens up another dimension to transport studies.   The attempt here is to study the transport characteristics of such a dusty plasma medium using the Molecular Dynamics (MD) simulations to essentially unfold the strong coupling and magnetize features discussed above.

 The quantitative assessment of any transport process is gleaned from how the mean square displacement of particles scales with time. In general, a random collisional process leads to a diffusive transport 
 \cite{chandrasekhar1943} and plays an important role in many physical, chemical, and biological systems \cite{difstudy1}.
  Mathematically, the diffusion coefficient is defined using the mean square displacement (MSD) of the particles, $\left<\Delta r^{2}\right>$ during time interval ($t$) as follows:
\begin{equation}
    D = \lim_{t\rightarrow \infty} \frac{\left< \Delta r^{2}\right>}{t}
\end{equation}
Equivalently it can also be expressed by the velocity auto-correlation function 
\begin{equation}
    D = \frac{1}{N}\sum_i \int_0^\infty \left< v_i(t).v_i(0) \right> dt
\end{equation}
Here, $v_i(t)$ and $v_i(0)$ correspond to velocity of the $i$-th dust particle at time $t$ and $0$, respectively. Here $N$ denotes the total number of dust particles in the system. The magnitude of the diffusion coefficient can typically be estimated by $D = l_{mfp}^2/t_{coll}$, where $l_{mfp}$ is the mean free path and $t_{coll}^{-1}$ is the collision frequency of the particles. For charged particles in a plasma, the presence of a magnetic field changes the effective values of these parameters. The transport along and perpendicular to the magnetic field   ($D_{\parallel}$ and $D_{\perp}$ respectively),  can have distinct values. The estimation of diffusion coefficients, their dependence on the external magnetic field, etc.,  for plasma medium have remained a topic of continued research for their relevance in the context of magnetic fusion and other areas \cite{bohm, TaylorMcnamara, PRL.60.1286, PRL.60.1290, vlad_vaul2002, Shakoori2022, LinQuasi2d, morfillnunomura2006, hou-shukla2009, ott_bonitz2014, hartman_donko2019, Dutta2021}. In the magnetized case, a simple estimate based on replacing the mean free step length with the gyroradius leads to a  $1/B^2$ scaling for the diffusion coefficient. It is, however,  not supported by experiments.  Taylor and McNamara have provided a theoretical description in support of $1/B$ scaling  \cite{TaylorMcnamara} of the diffusion coefficient.  

In 1988, a $1/B$ scaling was observed for a two-dimensional non-neutral pure electron plasma when the condition of   $r_L << \lambda_D$ ($r_L$ and $\lambda_D$, respectively, correspond to Larmor radius and Debye length) are met \cite{PRL.60.1286, PRL.60.1290}. 

The diffusive transport for dusty plasma has also been studied in various contexts. Simplest cases of self-diffusion by Brownian movement \cite{vlad_vaul2002} and diffusion by local electric field interactions \cite{Shakoori2022} were studied for weakly ionized complex dusty plasma. Additionally, diffusion in strongly coupled plasma has been studied both experimentally and numerically \cite{LinQuasi2d, morfillnunomura2006, hou-shukla2009} in the absence of a magnetic field. Simulations and experimental studies in the presence of a magnetic field have also been done earlier \cite{ott_bonitz2014, hartman_donko2019, Dutta2021} which show a  $1/B$ scaling with the magnetic field in specific parameter regimes. For instance, the simulation carried out by Ott et. al., \cite{ott_bonitz2014} has shown that for a one-component plasma and for small values of $\omega_{cd}/\omega_{pd}$ the diffusion coefficient does not show any variation with respect to the magnetic field. However,  $1/B$ scaling is observed at higher values of this parameter.

Our work aims here to look at the transport of charged dust particles (which can be looked upon as a one-component Yukawa system) across an applied magnetic field. At smaller values of $\omega_{cd}/\omega_{pd}$ we observe a pulsating dust cloud. The application of a collision term in fact recovers the $1/B$ scaling of the diffusion coefficient with respect to the magnetic field even for this particular regime. 
The open-source molecular dynamics simulations with the LAMMPS code have been utilized for this purpose. 
  We perform Molecular dynamics (MD) simulations for a collection of charged dust particles that are placed in the 2-D X-Y plane. 
The applied magnetic field is along the $\hat{z}$ direction. The dust particles get screened by the underlying plasma medium and interact with the screened-coulomb potential. The collisions with the neutral species in the plasma are considered to introduce the randomness in the charged dust particle trajectories.

The paper is organized in the following manner: In section \textsc{II}, we discuss the simulation geometry of the dusty plasma under consideration, along with simulation parameters and other details. In section \textsc{III},  the observations are analyzed to understand the dependence of diffusion coefficient on the applied magnetic field and the dust-neutral collision frequency.   We observe a $1/B$ variation of the diffusion coefficient for  $\nu \geq \omega_{cd}$ where $\omega_{cd}$ is the cyclotron frequency of the dust particle and $\nu$ is the dust neutral collision. 
Section \textsc{IV} summarizes the salient results. 

\label{intro}

\section{ MD Simulation Details}
\label{mdsim}

\begin{figure}
    \centering
    \includegraphics[width=0.96\linewidth]{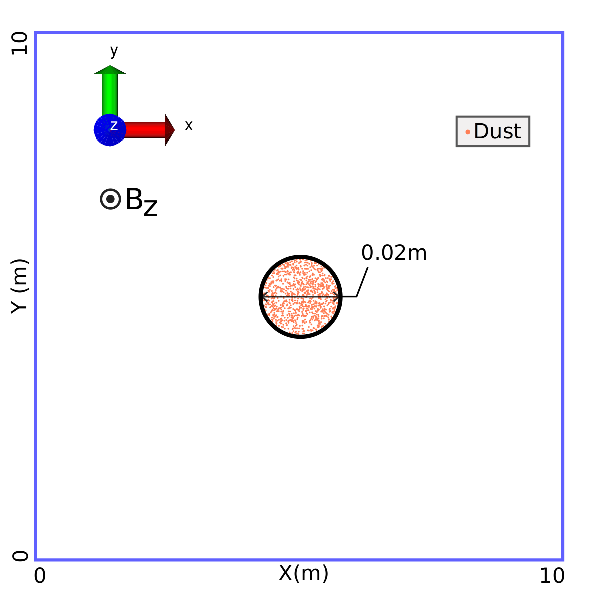}
    \caption{Schematic for 2D dusty plasma. At time t=0, a homogeneous magnetic field is applied along the z-axis. Particle dynamics lies in the x-y plane. }
    \label{fig:1}
\end{figure}

In our simulations, we consider a 2D geometry of dusty plasma with a magnetic field in the z-direction i.e., perpendicular to the plane of dynamics. The simulations are performed using the open-source classical MD simulation software LAMMPS \cite{LAMMPS}. A schematic of the simulation geometry is shown in FIG.\ref{fig:1}. The parameters (magnetic field, dust charge, and mass) chosen here correspond to nano-meter scale dust particles which are magnetized.  In simulations, we consider point particles. The typical parameters have been listed in table \ref{tab:table1}). 
\begin{table}
    \centering
    \begin{tabular}{c|c}
         Parameter&  Value\\
         \hline
         Mass, $m_d$& 8.1 $\times$ $10^{-17}$ kg\\
         Charge, $q_d$&100e\\
         No. of particles, N&1000\\
         Density, n&~ $10^7$ m$^{-2}$\\
         Temperature, $T_d$&132.45 K\\
         Time-period, $\omega_{pd}^{-1}$&0.3733 s$^{-1}$\\
         Time step, dt&$10^{-3} \omega_{pd}^{-1}$\\
         Runtime&$10^7$ steps\\
         Screening parameter, $\kappa$ (normalized)&0.5
         \end{tabular}
    \caption{Parameters as used in the simulations.}
    \label{tab:table1}
\end{table}

We consider a square simulation box of length $L_x = L_y = 10$m in both $\hat x$ and $\hat y$ directions with periodic boundary conditions. A circular region is chosen in the center of the box of radius $0.01m$ in which all $1000$ charged dust particles are randomly initialized. This results in an initial dusty plasma density ($n$) to be $\sim$ $10^{7} m^{-2}$. The mass of the dust particle is taken to be $8.1 \times 10^{-17}kg$ which is in agreement for particles of size $100-150$ $nm$. The charge ($q_d$) is taken to be $100e$ which is also in agreement with the charge acquired by nanometer-sized particles, where $e$ is the "electron charge". The coupling parameter, $\Gamma$, defined as the ratio of the potential energy to the kinetic energy of the dust particles ($\Gamma = q_{d}^{2}/r k_B T_d \exp{-r/\lambda_d}$), is taken to be $10$ which lies well in the strong coupling regime. Here, $q_d$ is the dust charge, $r$ is the inter-particle distance, and $T_d$ is the dust temperature, which, for $\Gamma = 10$ comes out to be $132.45K$. The timescale is normalized by factor $\omega_{pd}^{-1}$, where $\omega_{pd}=(nq_d^2/m_d\epsilon_{o})^{1/2}$ is the dust-plasma frequency. 

 A Yukawa one-component plasma potential  is considered for the inter-dust interaction potential 
 \begin{equation}
     \varphi(r) = \frac{q_d}{r} \exp{(-\kappa r)}
     \label{eq1}
 \end{equation}
 Here $\kappa$ is the inverse of the shielding length of the dust particles by the background plasma.  It is related to  Debye shielding by the relationship  $\kappa = \lambda_d^{-1}$. 

 We choose the normalized value of the shielding parameter $\kappa =0.5$ with $a$ (inter-dust distance chosen initially) as the normalization constant. 

 A cut-off scale for interaction is chosen to be  $r_c =8a$, where $a$ is given as $1/\sqrt{n\pi}$.  The system is simulated under thermal equilibrium conditions by setting  Nose-Hoover thermostats \cite{nose1984unified,hoover1985canonical}. 
 
 The dust particle locations are evolved using Newton's equation of motion which is written as follows.
\begin{equation}
     m\mathbf{\ddot{R}} = q_d (-\nabla \varphi + \mathbf{\dot{R}\times \mathbf{B}}) - m\nu\mathbf{\dot{R}}
     \label{eq3}
 \end{equation}

 Here $\mathbf{R}$ is the location of the dust particle, $-\nabla \varphi$ is the electric field generated by the Yukawa potential, and $\mathbf{B}$ is the magnetic field. We have also considered collisional damping due to neutral species in the plasma in the dust equation of motion. This is depicted by $\nu$ in the equation. 

 The above equation is simulated using the open source code LAMMPS \cite{LAMMPS}. The collisional damping is taken care by the Langevin damping parameter while using the Langevin thermostat \cite{schneider1978molecular}. The velocity equations are integrated using the Stoermer-Verlet time integration algorithm (velocity-Verlet) \cite{verlet}. 
  
 The system is configured to evolve with a timestep of $dt = 10^{-3}\omega_{pd}$. This resolves the dust plasma period. The time integration is followed for more than $10$ million timesteps. The numerical estimates for the diffusion coefficient are then obtained and their dependence on the magnetic field and the collision frequency is ascertained. The observations are summarized in the next section. 
 
\section{Diffusive transport of dust particles}
\subsection{The case of no dissipation $\nu = 0$ }
The first consider the case when the dissipation due to the dust neutral collision is absent, so as to have $\nu = 0$. Here the dust particle evolution is governed by the self-consistent electric field and the applied magnetic field. The self-consistent electric field has a repulsive character and it causes the dust particles to move away from each other in the $x-y$ plane. The application of an external magnetic field along $\hat{z}$  has a confining role. The combination is expected to induce  $\mathbf{E} \times \mathbf{B}$ drift. This drift will be in the $\theta$ direction for the radially symmetric dust charge distribution. This ensures that the dust particles remain confined.  A comparison of the spatial distribution of the dust particle distribution in the absence as well as in the presence of a magnetic field can be observed in Fig.2 (a) and (b) respectively at the same time. It can be observed that the spatial distribution of the dust has a greater radial extent in the absence of the magnetic field.

\begin{figure}
 \includegraphics[width=\linewidth]{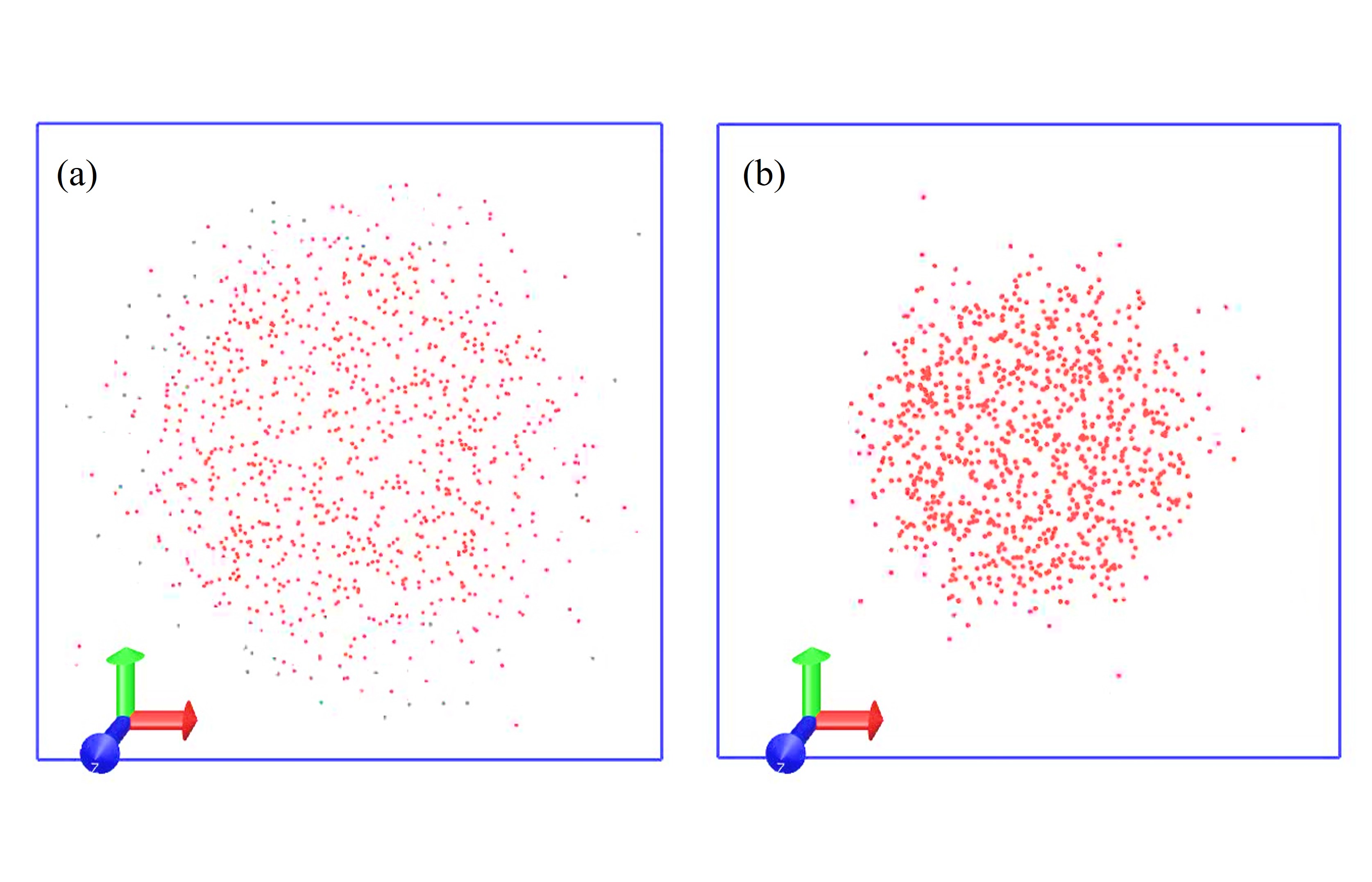}
     \caption{Distribution of dust particles in simulation box after $t \sim 1000\omega_{pd}^{-1}$. In subplot (a) the magnetic field is absent, whereas in subplot (b) a typical magnetic field of 0.035T is applied. The particles travel more in the absence of a magnetic field as compared to when an external magnetic field is applied.}
  \label{fig:2}
 \end{figure}

\begin{figure}
    \centering
    \includegraphics[width=0.96\linewidth]{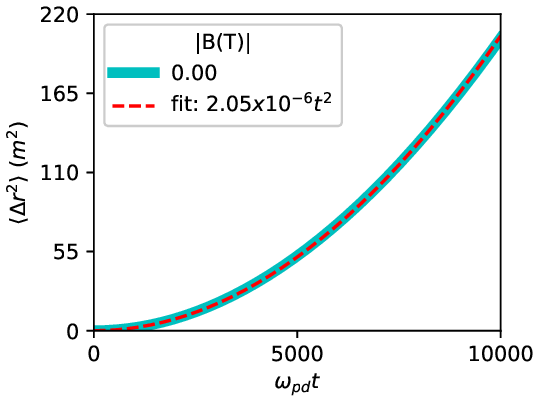}
    \caption{Mean square displacement (MSD) plotted against normalized time for the unmagnetized case showing a ballistic transport.}
    \label{fig:3}
\end{figure}

 Fig. \ref{fig:2}.
 For the unmagnetized case, the mean square displacement (MSD) of the dust particles from their original position displays a $t^2$ scaling with time as shown in Fig.\ref{fig:3}. Thus the expansion is ballistic in nature. 

\begin{figure*}
    \centering
    \includegraphics[width=0.9\linewidth]{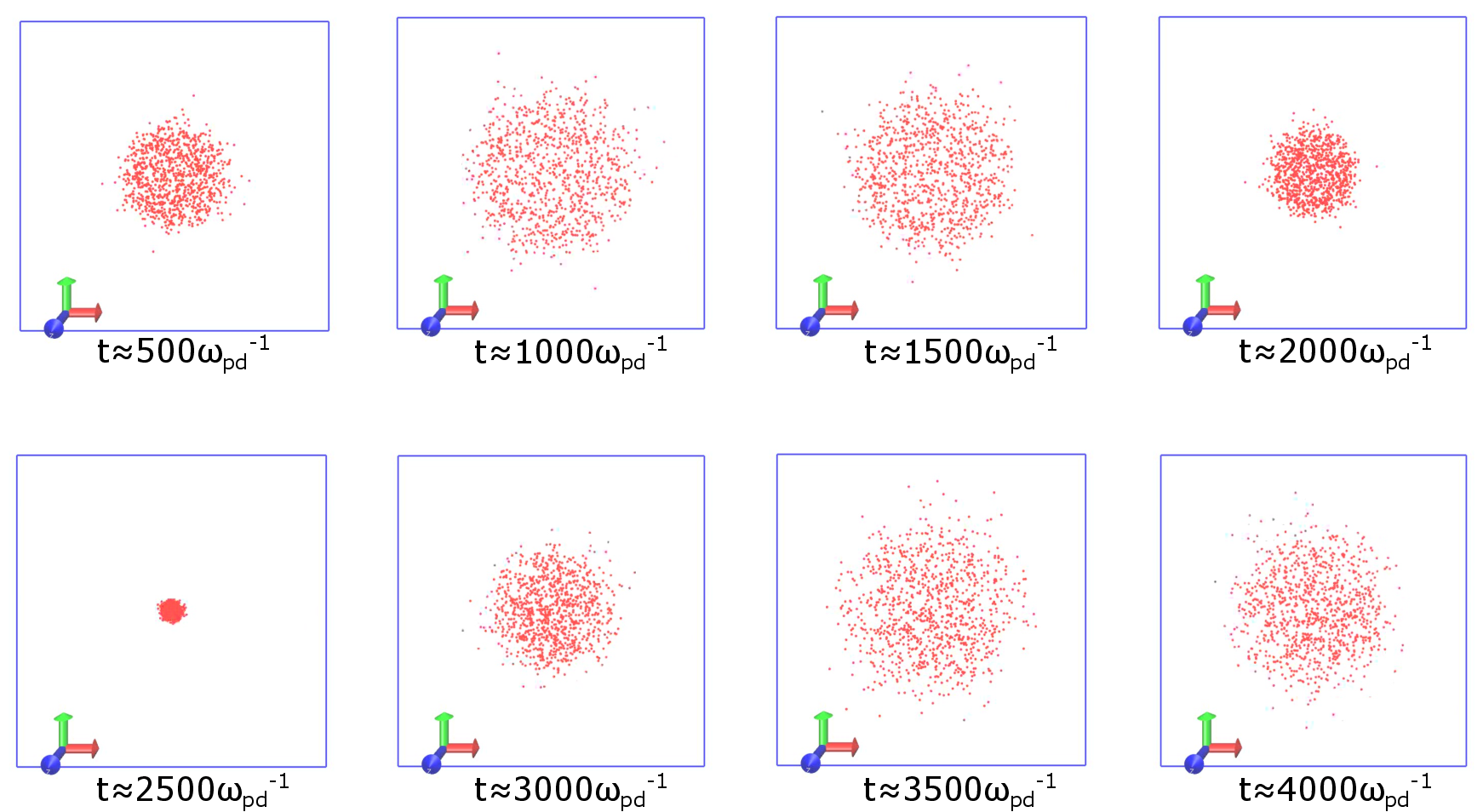}
    \caption{Time evolution of dust particles in the presence of a perpendicular magnetic field of 0.035T.}
    \label{fig:4}
\end{figure*}
\begin{figure*}
    \centering
    \includegraphics[width=0.9\linewidth]{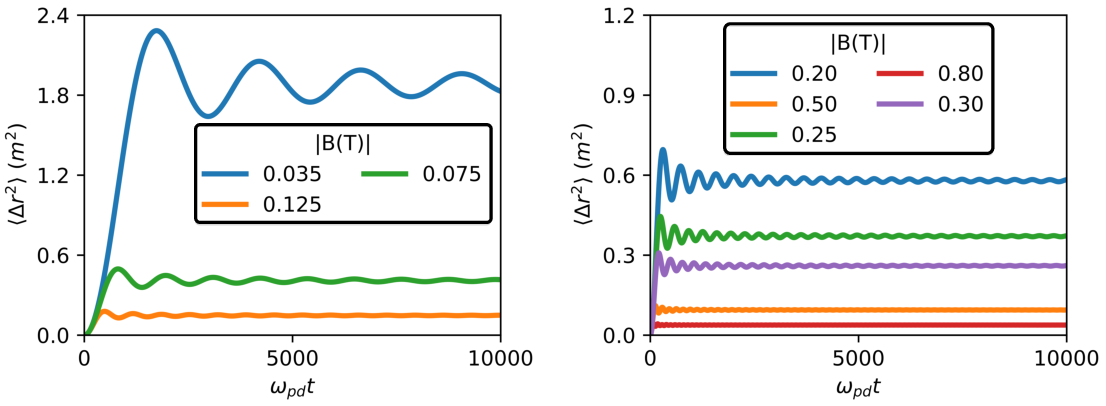}
    \caption{MSD values for different magnetic field values.}
    \label{fig:5}
\end{figure*}

 In the presence of a magnetic field, however, the spatial distribution of dust shows radial pulsations.  In Fig. \ref{fig:4} the spatial distribution of the dust has been shown at various times. The pulsations are evident as the radial extent can be first seen to increase (e.g. $t = 500$ to $t = 1500$), thereafter it gets reduced ($t = 2000$) and squeezed further at ($t = 2500$). It is then followed up by the expansion phase. This cycle continues and can be observed more clearly in the plot of mean square displacement (MSD) as a function of time in Fig. \ref{fig:5}
 for various values of the magnetic field. A  movie showing the dust evolution illustrating the phenomena of pulsations has been attached as supplementary material  [].
It can be seen from Fig. \ref{fig:5} that oscillations are around some average value of MSD. This average value is high for weak magnetic fields. 
\begin{figure}
    \centering
    \includegraphics[width=0.9\linewidth]{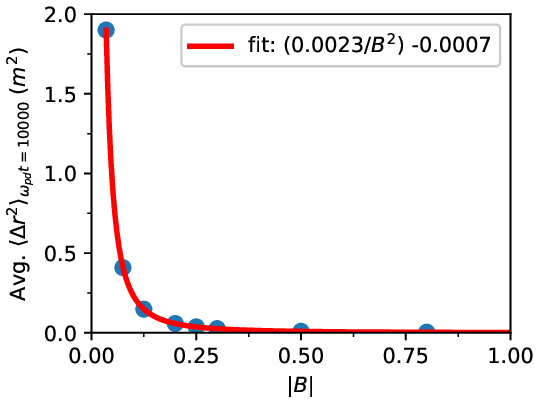}
    \caption{Averaged out MSD after final timestep plotted for different values of the magnetic field.}
    \label{fig:6}
\end{figure}
\begin{figure}
    \centering
    \includegraphics[width=0.9\linewidth]{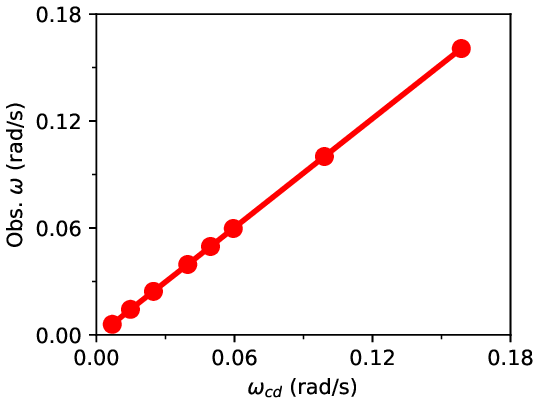}
    \caption{Pulsation frequency observed v/s actual gyrofrequency for different magnetic field values. It can be easily seen that pulsation happens around the gyrofrequency.}
    \label{fig:7}
\end{figure}
The dependence of the average MSD as a function of the magnetic field has been shown in Fig. \ref{fig:6}. The fit clearly indicates a clear $1/B^2$ dependence on the magnetic field. We also observe that the oscillation frequency increases with the strength of the magnetic field. In fact, the plot of the observed oscillation frequency shown in Fig. \ref{fig:7} with respect to the dust gyrofrequency shows that the oscillations are essentially the same as the dust gyro-frequency.  

Let us now try to understand these features. The dust particles while rotating at the gyrofrequency can have displacement which can have a maximum value of twice the gyro-radii. It should be noted that our simulations have been carried out with a thermostat of fixed temperature. Thus the velocity of the dust particles is of the order of typical thermal velocity which is independent of the magnetic field. The 
$(\Delta r)^2 \propto (v_{th}/\omega_{cd})^2 \sim 1/B^2$. This explains the observed scaling.

\subsection{Dust transport in the presence of finite dust neutral collision frequency $\nu$.}
\begin{figure*}
    \centering
    \includegraphics[width=0.9\linewidth]{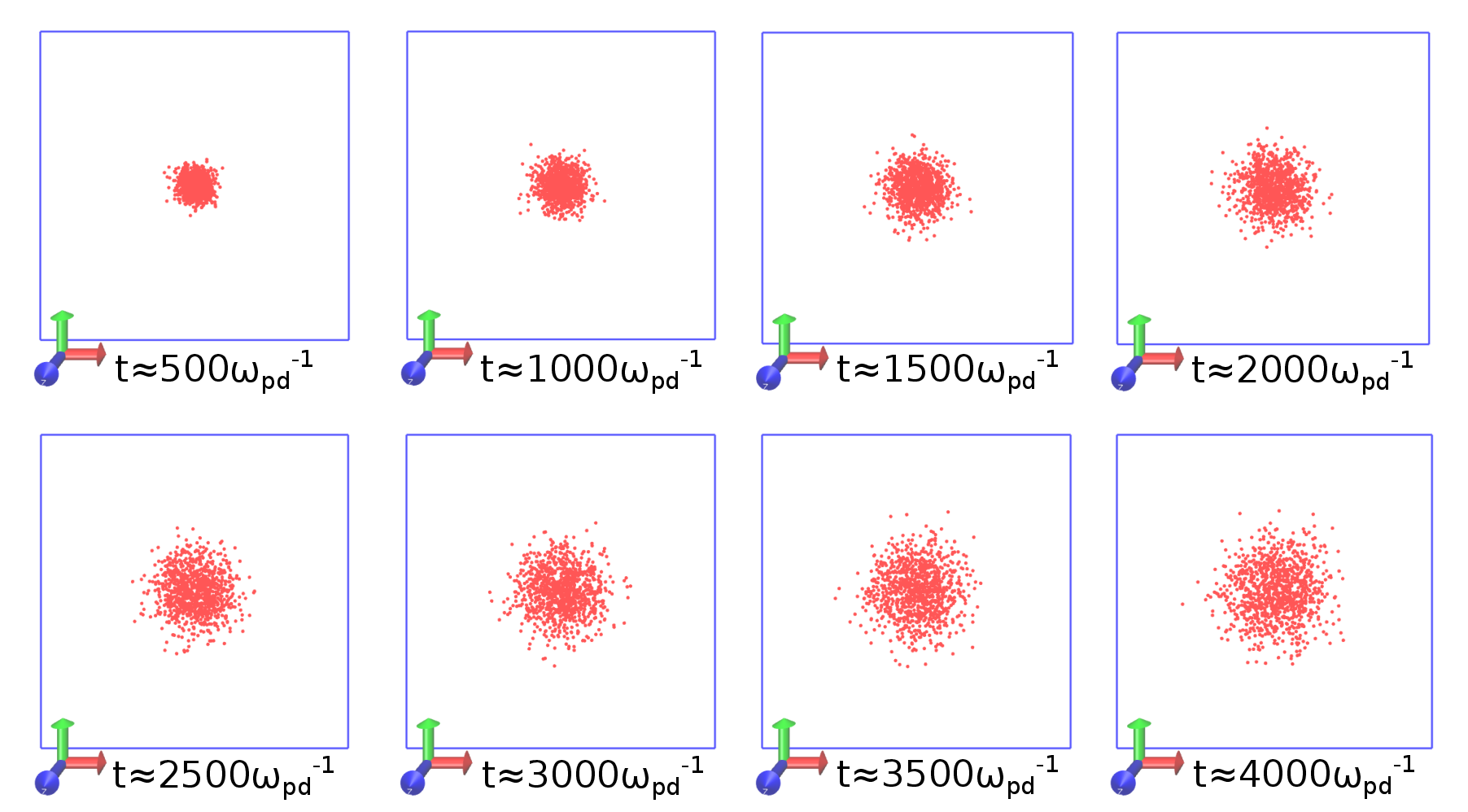}
    \caption{Time evolution of dust particles colliding at 0.1 collisions per second frequency in the presence of a perpendicular magnetic field of 0.035T and $\nu=0.1$.}
    \label{fig:8}
\end{figure*}
The radial pulsations observed when $\nu = 0$,  disappear when $\nu$ is chosen to be finite. The finite value of $\nu$ can be viewed as dissipation arising from the dust-neutral collision. The dust particles while they are exhibiting gyro-oscillations may encounter collisions with the neutral particles which randomize the trajectories.   Fig. \ref{fig:8}, shows the snapshots at various times of the spatial distribution of the dust particles in the $X-Y plane$. Unlike Fig. \ref{fig:4} there are now no radial pulsations. Instead, the radial extent of the dust particles keeps increasing. The magnetic field for this case has a value of  $0.035T$ and the collision frequency $\nu = 0.1$. 
\begin{figure}
    \centering
    \includegraphics[width=0.9\linewidth]{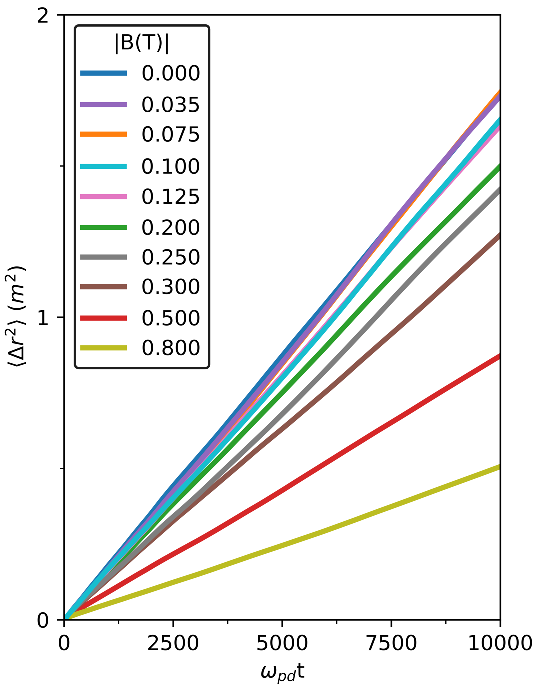}
    \caption{The “Mean Square Displacement (MSD)” of dust particles as a function of time has been shown (left) for various values of the magnetic field B. The straight lines indicate that the transport is diffusive. The coupling parameters for this case correspond to a value of 10 and $\nu = 0.1$}
    \label{fig:9}
\end{figure}

\begin{figure}
    \centering
    \includegraphics[width=0.9\linewidth]{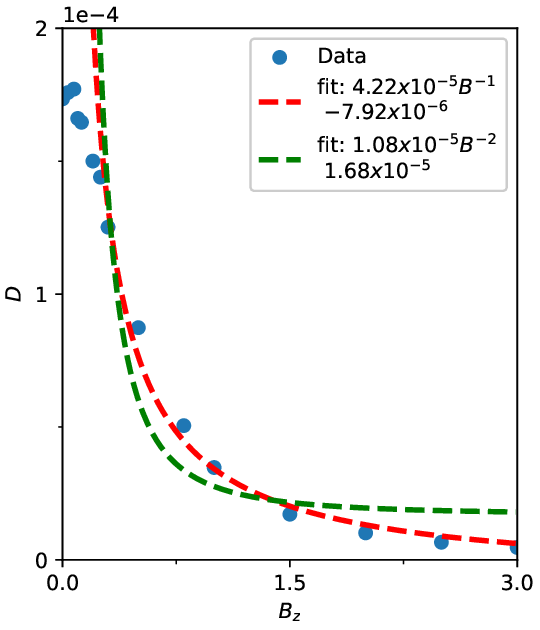}
    \caption{ The diffusion coefficient D as a function of the magnetic field has been plotted (right), showing $1/B$ scaling. The coupling parameters for this case correspond to a value of 10 and $\nu = 0.1$}
    \label{fig:10}
\end{figure}
The MSD of the dust particles, in this case, increases linearly with time. This has been shown in the plot of Fig. \ref{fig:9} where the MSD as a function of time for various values of the magnetic field has been shown. The plots clearly show a linear behaviour with time indicating a diffusive process at work. The scaling of the diffusion coefficient with magnetic field evaluated from this data has been shown in Fig. \ref{fig:10}. The simulation observations are shown as blue-colored dots. It can be observed that the blue colored dots closely fit the curve having a $1/B$ scaling with the magnetic field. The $1/B^2$
The scaling curve depicted by the green dashed line does not show agreement with the results. It is also observed that at low values of the magnetic field, there are deviations from even the $1/B $ scaling. This scaling, therefore, is an illustration of the magnetized behavior of the dust dynamics. 

The $1/B$ scaling of the diffusion coefficient has been analytically shown by Taylor and McNamara \cite{TaylorMcnamara} for the coulomb plasma. Here too we observe the same scaling, even though the dust particles are interacting via screened coulomb potential. This can be understood by realizing that the screening merely restricts the distance of interparticle interaction and has no magnetic field dependence. 

We have also carried out simulation studies to understand the behavior of the transport coefficient 
as a function of $\nu$. The behavior of the diffusion coefficient as a function of $\nu$ is non-monotonic. We observe two different scalings depending on whether $\nu < \omega_{cd}$ or $\nu > \omega_{cd}$.

\begin{figure}
    \centering
    \includegraphics[width=0.95\linewidth]{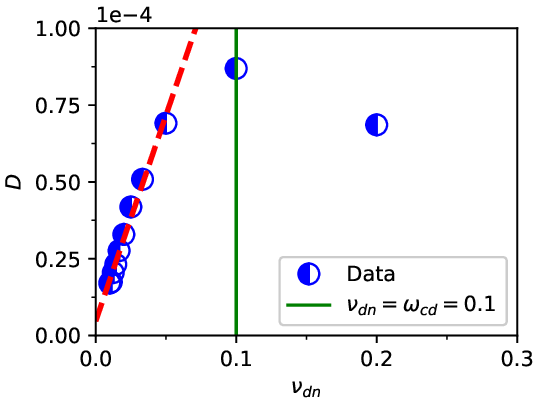}
    \caption{The diffusion coefficient $D$ as a function of the collision frequency has been plotted at $B_z=0.5T$. We observe two regions of different dependency on $\nu_{dn}$. Here, we see a $\propto \nu_{dn}$ scaling for $\nu < \omega_{cd}$.}
    \label{fig:nuvarfit}
\end{figure}
\begin{figure}
    \centering
    \includegraphics[width=0.95\linewidth]{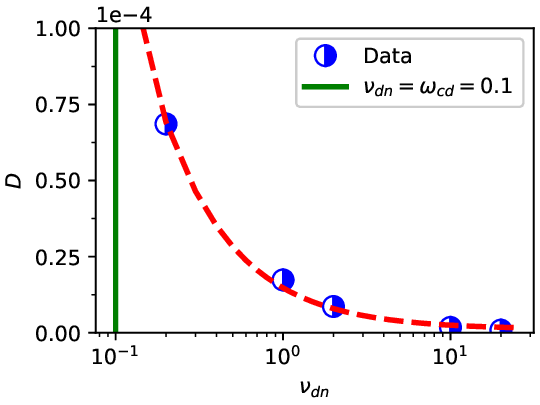}
    \caption{Here, a scaling of $1/\nu_{dn}$ is observed for $\nu_{dn} > \omega_{cd}$ for the diffusion coefficient $D$ plotted against the collision frequency at $B_z=0.5T$. }
    \label{fig:nuvarfit}
\end{figure}
It can be observed from Fig.11 that for $\nu_{dn} < \omega_{cd}$ the value of the diffusion coefficient increases linearly with the frequency. The green vertical line has been put to show the value of  $\nu_{dn}$ which matches with $\omega_{cd}$. 

The linear scaling with collision frequency at low magnetic field is understood by realising that the step length in this case is limited by the collision mean free path. 
In the other limit for which $\nu_{dn} > \omega_{cd}$ an inverse dependence with collision frequency is observed as shown in Fig.12. This happens as the presence of a strong magnetic field restricts the step length to the gyroradii. 

The other parameter in the system is the screening distance $1/\kappa$. The transport studies were also carried out to study the dependence of $D$ on $\kappa$. It is observed that the value of $D$ typically reduces with $\kappa$ when its value is lower than the inverse of the interparticle distance $a$. However, with an increasing value of $\kappa$, there are oscillations in $D$ indicating 
a tendency towards saturating asymptotically. This can be understood by realising that for high values of $\kappa$ shielding is perfect and within the interparticle spacing. Thus the interaction amongst dust particles is extremely weak in this regime.

\begin{figure}
    \centering
    \includegraphics[width=0.9\linewidth]{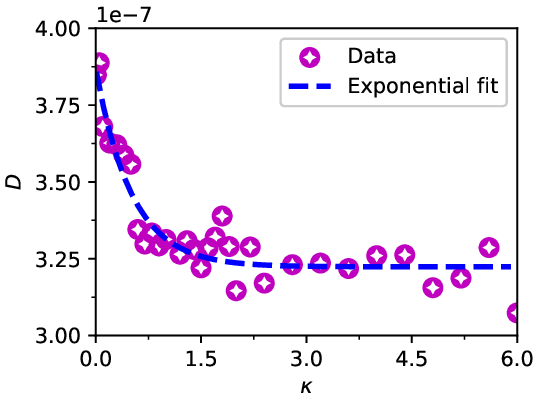}
    \caption{The diffusion coefficient $D$ as a function of the shielding parameter $\kappa$ has been plotted. The magnetic field value is $0.5T$ and the collision frequency is $0.1$.}
    \label{fig:kappavar}
\end{figure}

\section{Summary}
We have investigated the transport of strongly coupled dust particles in a magnetized dusty plasma in a two-dimensional plane. Molecular Dynamics simulations have been performed for this purpose.  The dust particles are essentially massive high charged particles interspersed in a background plasma. The screening due to background plasma leads to Yukawa interaction amongst these particles. The externally applied magnetic field is uniform and directed normally to the plane.

The dust particle cloud is observed to exhibit radial pulsation at the gyrofrequency. However, when a collisional term is introduced (which can mimic the dust-neutral collision) the dust cloud expands showing radial diffusion. The diffusion coefficient is found to scale as $1/B$. Such a scaling has also been observed in the context of electron-ion plasma. The theoretical interpretation of the same has been provided in the paper by Taylor and McNamara\cite{TaylorMcnamara} The fact that the interaction potential is Yukawa here instead of pure Coulomb for the electron-ion plasma does not seem to affect this scaling with magnetic field.

We have also investigated the scaling of the diffusion coefficient with respect to the collision frequency. The scaling is observed to be linear when the collision frequency is less than the gyrofrequency and the plasma is magnetized. While for the unmagnetized case (when collision frequency exceeds the gyroradius the inverse dependence on i is observed as expected for a single particle \cite{braginskii1965transport}.

Dusty plasma is being extensively studied in laboratories and has applications in many contexts.  In particular, the magnetized device like a magnetron sputtering system which uses a cathode target for the deposition of crystalline thin films in the presence of a magnetic field often encounters charged dust particle clustering. They may also form crystalline structures wherein tens to hundreds of nanometer-sized grains are clustered together\cite{nisha2024overall}.  The transport and dispersal of such dust clouds are often crucial for better operating conditions.  The work here thus has significance for laboratory plasma like magnetron sputtering used for the deposition of thin films. The growth of crystalline microstructure or nanostructure is a critical requirement for numerous applications in nanoscience. The experimental work on magnetron devices is in progress and will be presented in a future publication.

 \section*{Acknowledgements}
AD acknowledges the support from  Core Research Grant No. CRG/2022/002782  of the Department of Science and Technology (DST), and the  J.C. Bose
Fellowship grant (Grant No. JCB-000055/2017) for this work. BBS acknowledges support from CRG/2022/001470 of DST. The authors thank the IIT Delhi HPC facility for computational resources. ASK thanks the Council for Scientific and Industrial Research (Grant no. 09/0086(13737)/2022-EMR-I) for financial assistance. ASK would also like to thank Mamta Yadav and Sooryansh Asthana for discussions related to LAMMPS and theoretical modeling respectively.
   
\bibliography{ref}
\end{document}